\documentclass[12pt,a4paper]{article}

\usepackage{hyperref}

\usepackage{graphicx}
\usepackage{color}

\usepackage{amssymb}
\usepackage{gensymb}
\usepackage{textcomp}
\usepackage[version=4]{mhchem}

\usepackage{setspace}

\title{First-principles LCAO study of the low and room temperature phases of CdPS$_3$}

\author{Alexei Kuzmin\\ \\  \small\em Institute of Solid State Physics, University of Latvia,\\ \small\em 8 Kengaraga street, Riga LV-1063, Latvia\\ \small E-mail: a.kuzmin@cfi.lu.lv}

\date{Received \today}

\begin{document}
	
\maketitle

\newpage

\begin{abstract}
The electronic and atomic structure of a bulk 2D layered van-der-Waals compound CdPS$_3$ was studied
in the low ($R3$) and room ($C2/m$) temperature phases using first-principles calculations within the periodic linear combination of atomic orbitals method with hybrid meta exchange-correlation M06 functional. The calculation results reproduce well the experimental crystallographic parameters.
The value of the indirect band gap $E_g$=3.4~eV for the room-temperature monoclinic $C2/m$ phase is close to the experimental one, while the indirect band gap $E_g$=3.3~eV was predicted for the low-temperature trigonal $R3$ phase. The effect of hydrostatic pressure on the band gap in both phases was studied in the pressure range from 0 to 40~GPa. In both cases, the pressure dependence of the band gap passes through a maximum, but at different pressures. In the $R3$ phase, the band gap reaches its maximum value of $\sim$4~eV at $\sim$30~GPa, whereas in the $C2/m$ phase, the maximum value of $\sim$3.6~eV is reached already at $\sim$8~GPa. 
  
\end{abstract}

\newpage

\section{Introduction}\label{s:intro}

Two-dimensional (2D) layered compounds represent a rich class of functional materials with tunable electronic, magnetic  and optical properties \cite{Gupta2015,Lv2015,Xiao2019}. 
Their atomic structure is distinguished by strong in-plane bonding and weak van-der-Waals (vdW) out-of-plane interactions, which make it possible to exfoliate the material into 2D (single-)layers and  pack them into heterostructures \cite{Liu2016,THANH2018}.  More than a few hundred 2D layered materials are known today, while more than five thousand such compounds were predicted by high-throughput  first-principles calculations \cite{Mounet2018}. 

\begin{figure}[t]
	\centering
	\includegraphics[width=0.7\textwidth]{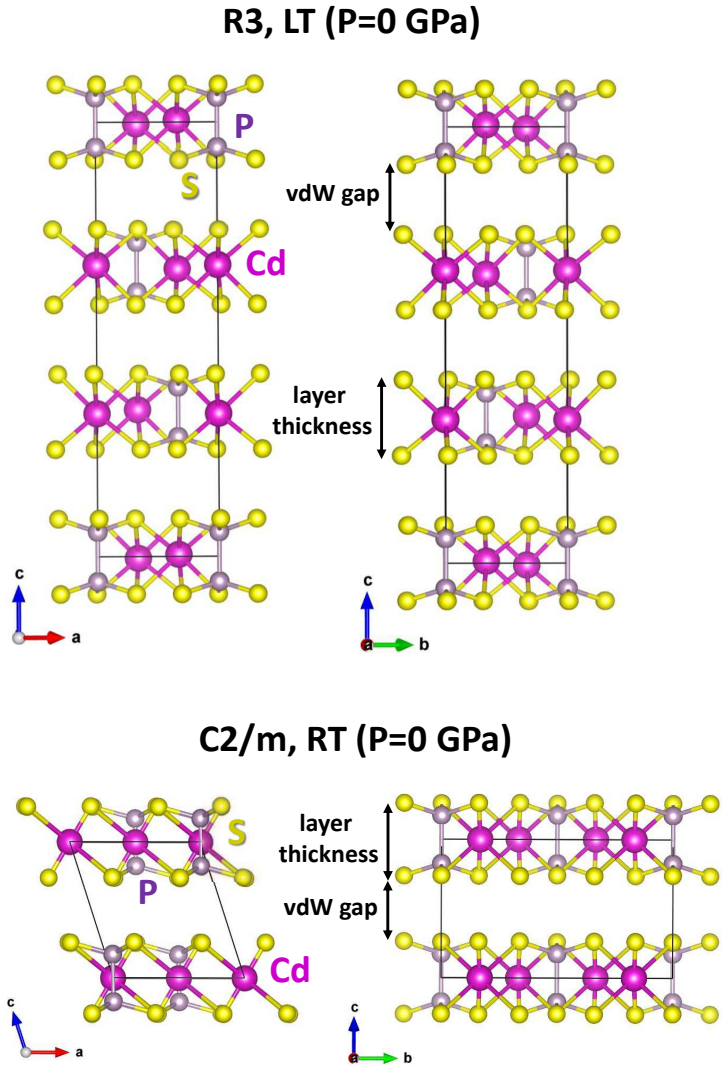}
	\caption{ Crystallographic structure  of CdPS$_3$ in the low-temperature (LT) trigonal $R3$ \protect\cite{Boucher1995} and 
	room-temperature (RT) monoclinic $C2/m$ \protect\cite{Klingen1973,Ouvrard1985}  phases at pressure $P=0$~GPa. The interlayer spacings (vdW gaps) between neighbouring layers and the layer thicknesses are indicated. The illustrations were created using the VESTA software \protect\cite{VESTA}.}
	\label{fig1}
\end{figure}

Among different groups of 2D layered materials, metal thio(seleno)phosphates (MTPs), being  moderate- to wide-bandgap semiconductors, represent a particular interest \cite{Susner2017}.
Their crystallographic structure (Fig.\ \ref{fig1}) consists of the [P$_2$S(Se)$_6$]$^{4-}$ anion framework forming slabs, staking along the $c$-axis and stabilized by metal cations, which can be group I and II elements, most of the transition metals, some heavier metals as well as a few lanthanides and actinides \cite{Susner2017}. 
It is the type of metal cation that is responsible for a wide variety of the MTP functionalities.
While MTP materials are currently a hot topic of research, some of them were less studied.
Here the case of CdPS$_3$ will be discussed.

At room temperature, CdPS$_3$ crystallizes in the monoclinic $C2/m$ (No. 12) phase with two formula units in the primitive unit cell, but with four formula units in the crystallographic unit cell \cite{Klingen1973,Ouvrard1985,BOUCHER1994}. 
The atoms occupy the following Wyckoff positions: Cd 4g(0,y,0), P 4i(x,0,z), S1 4i(x,0,z), S2 8j(x,y,z) (Table\ \ref{table1}). The experimental band gap $E_g$ is equal to 2.95~eV \cite{Fuentealba2020},  3.06~eV \cite{Calareso1997} and 3.5~eV \cite{Brec1979,YANG2002,Du2016}, which makes it possible to classify CdPS$_3$ as a wide-bandgap semiconductor.

When the temperature drops below 228~K, CdPS$_3$ transforms into the low-temperature trigonal $R3$ (No. 146) phase with two formula units in the primitive unit cell, but with six formula units in the crystallographic unit cell \cite{LIFSHITZ1983,Boucher1995,Sourisseau1996}.
The atoms are located at the following Wyckoff positions: Cd1 3a(0,0,z), Cd2 3a(1/3,2/3,z), P1 3a(2/3,1/3,z), P2 3a(2/3,1/3,z), S1 9b(x,y,z), S2 9b(x,y,z) (Table\ \ref{table1}). To our knowledge, the experimental value of the band gap for the $R3$ phase has not been reported.

The layered structure of CdPS$_3$ can also be characterized by the interlayer spacings (vdW gaps) 
and the layer thickness, defined as the distance between the planes containing sulfur atoms on opposite sides of the layer  (Fig.\ \ref{fig1}). 
The experimental interlayer spacings and the layer thicknesses are close in both phases and 
are, respectively, equal to  $\sim$3.12~\AA\ and $\sim$3.37~\AA\ for the $R3$ phase \cite{Boucher1995} and to 
$\sim$3.16~\AA\  and $\sim$3.38~\AA\ for the $C2/m$ phase  \cite{Ouvrard1985,BOUCHER1994}. 

Most recent theoretical studies of CdPS$_3$ have used the plane-wave DFT approach \cite{Liu2014,Xiang2016,Hashemi2017}.

The band gap $E_g$ equal to 2.96~eV was obtained for bulk CdPS$_3$ and 3.03~eV for its single-layer in  \cite{Liu2014} using a hybrid HSE06 exchange-correlation functional \cite{HSE06}. It was concluded that the electronic structure of single-layer CdPS$_3$ is suitable for visible-light driven photocatalysis, and the position of its band edges relative  to the water redox potential makes single-layer CdPS$_3$ a promising candidate for photocatalytic splitting of water \cite{Liu2014}. 

Vibrational properties of MTPs were studied in \cite{Hashemi2017}
using the generalized gradient approximation (GGA) for the exchange-correlation energy 
in the form of the PBE functional \cite{PBE}. While the structure of CdPS$_3$ was satisfactory reproduced, the calculated band gap value of 1.93~eV was underestimated compared to the known experimental values. 
The calculated Raman active modes were analysed in detail for the unstrained and strained layers, and the frequency shifts of the modes in response to biaxial strain were determined  \cite{Hashemi2017}. The computed  elastic properties indicated that MTPs (including CdPS$_3$) have very soft lattice compared to other layered materials (e.g., MoS$_2$ and graphene)  \cite{Hashemi2017}.

The effect of strain on the electronic structure of MPX$_3$ (M = Zn, Cd; X = S, Se) single-layers
was also studied in \cite{Xiang2016}. It was found that CdPS$_3$ always has indirect band gap
under the strain load of up to $\pm$10\%. Moreover, under the compressive strain, the band gaps of MPX$_3$ single-layers increase firstly and then decrease, whereas the tensile strain leads only to a decrease of the band gaps. It was concluded that the MPX$_3$ single-layers are promising candidates for the tunable electronic structures by strain engineering \cite{Xiang2016}.

In this study, the electronic structure of the low ($R3$) and room ($C2/m$) temperature phases of bulk CdPS$_3$ was studied using first-principles linear combination of atomic orbitals (LCAO) method, and the effect of hydrostatic pressure on the band gap in both phases was evaluated.

\section{Methodology}\label{s:method}

First-principles calculations of  CdPS$_3$ were performed by the linear combination of atomic orbitals (LCAO) method using the CRYSTAL17 code \cite{crystal17}.
The basis sets for P, S and Cd atoms were chosen in the form of all-electron triple-zeta valence (TZV) basis sets augmented by one set of polarization functions (pob-TZVP) \cite{Oliveira2019}.

The evaluation of the Coulomb and exchange series was done with the accuracy controlled by a set of tolerances, which were selected to be (10$^{-8}$, 10$^{-8}$, 10$^{-8}$, 10$^{-8}$, 10$^{-16}$). The Monkhorst-Pack scheme \cite{Monkhorst1976} for an 8$\times$8$\times$8 $\textbf{k}$-point mesh was used to integrate the Brillouin zone.  The SCF calculations were performed employing the M06 \cite{M06}  functional  with a 10$^{-10}$ tolerance for the total energy change.

The lattice parameters and atomic fractional coordinates were optimized for 
the low-temperature (LT) trigonal $R3$ \cite{Boucher1995} and 
room-temperature (RT) monoclinic $C2/m$ \cite{Ouvrard1985,BOUCHER1994} CdPS$_3$ phases. 
The atomic charges were estimated from the Mulliken population analysis \cite{Mulliken1955}. 
The calculated lattice parameters ($a$, $b$, $c$, $\beta$), atomic fractional coordinates  ($x$, $y$, $z$) and the values of the band gap ($E_g$)  are reported in Table\ \ref{table1}. 

\begin{figure}[t]
	\centering
	\includegraphics[width=0.7\textwidth]{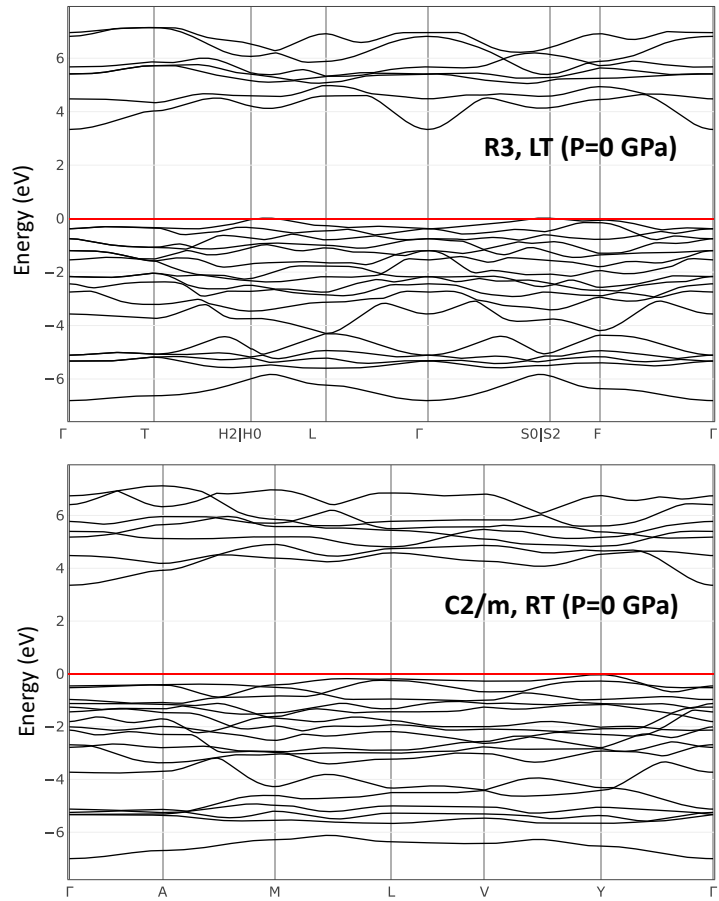}
	\caption{ Band structure diagrams for the low-temperature (LT) trigonal $R3$ and room-temperature (RT) monoclinic $C2/m$  CdPS$_3$ phases at 0~GPa. The energy zero is set at the top of the valence band (Fermi energy position). }
	\label{fig2}
\end{figure}

\begin{figure}[t]
	\centering
	\includegraphics[width=0.95\textwidth]{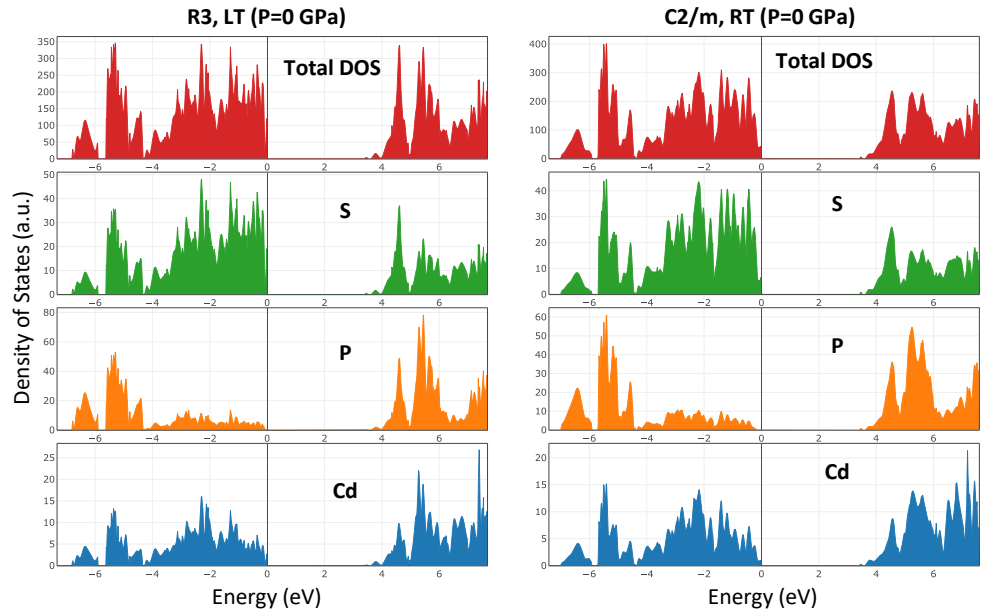}
	\caption{ Total and projected onto atoms density of states (DOS) for the low-temperature (LT) trigonal $R3$ and room-temperature (RT) monoclinic $C2/m$  CdPS$_3$ phases at 0~GPa. The energy zero is set at the top of the valence band (Fermi energy position). }
	\label{fig3}
\end{figure}

\begin{figure}[t]
	\centering
	\includegraphics[width=0.95\textwidth]{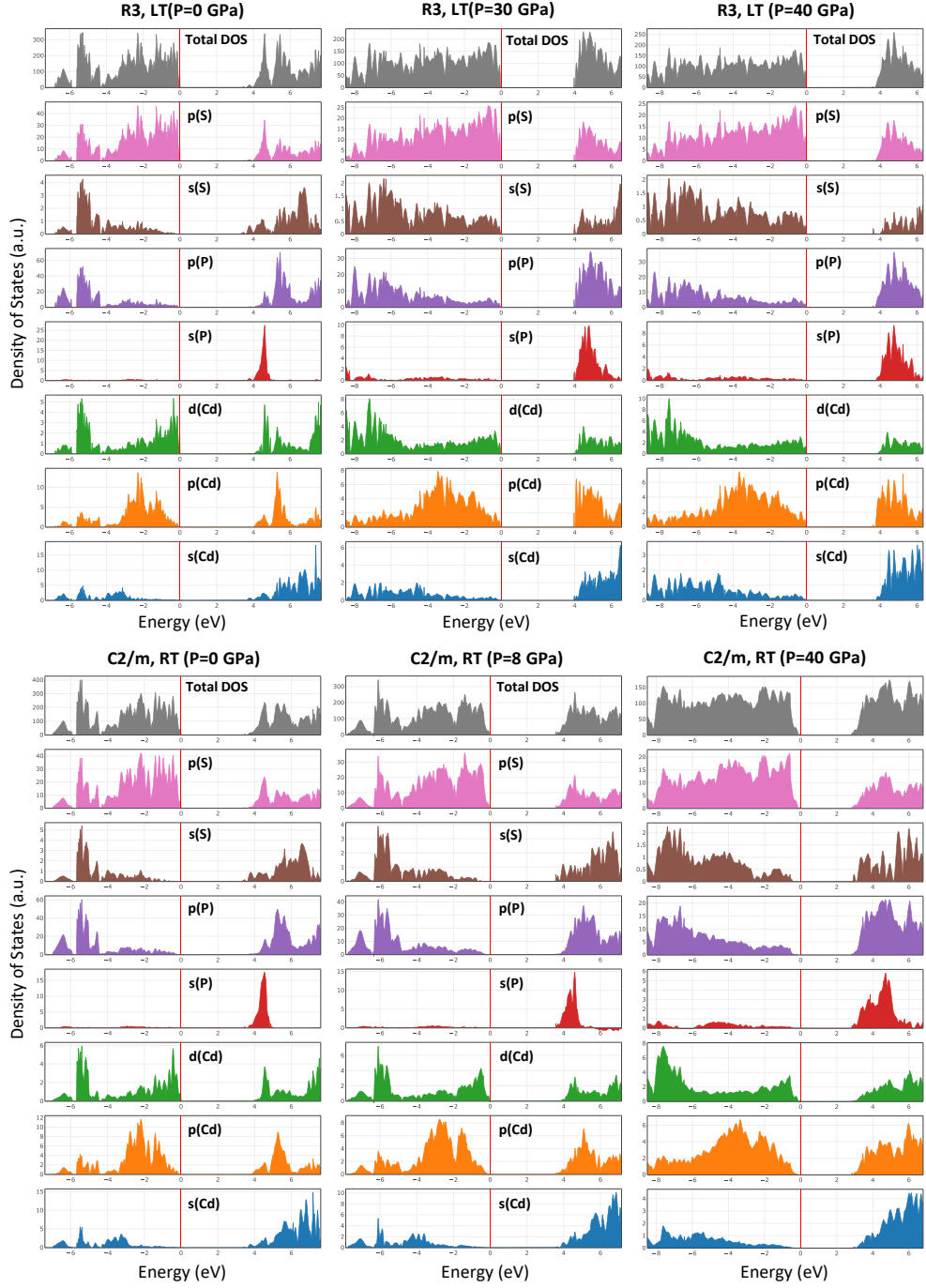}
	\caption{Total and projected onto the set of atomic orbitals density of states (DOS) in CdPS$_3$ for the low-temperature (LT) trigonal $R3$ phase at 0, 30 and 40 GPa and room-temperature (RT) monoclinic $C2/m$ phases at 0, 8 and 40~GPa. The energy zero is set at the top of the valence band (Fermi energy position). }
	\label{fig4}
\end{figure}

Calculated  band structures and total/projected density of states (DOS)  are shown in Figs.\ \ref{fig2}, \ref{fig3} and \ref{fig4}. The band structures were calculated along band paths in the Brillouin zone chosen according to \cite{Hinuma2017}.
Note that in the rhombohedral space group $R3$, belonging to the trigonal crystal system, two
topologically different shapes of the Brillouin zone (and, respectively, sets of the high-symmetry points) are possible depending on the ratio between $a$ and $c$ lattice parameters \cite{Hinuma2017,Aroyo2014}. In our case $\sqrt{3}a < \sqrt{2}c$ (Table\ \ref{table1}), and
the Brillouin zone has the topology of the truncated octahedron \cite{Aroyo2014}.       

Additionally to structural parameters, the atomic displacement parameters (ADPs) $B_{\rm eq}$ were determined for the $R3$ phase at 133~K and for the $C2/m$ phase at 298~K in order to compare them with the available experimental data (Table\ \ref{table1}).  
ADPs were calculated  in the harmonic approximation for the supercell size 2$\times$2$\times$1 from the eigenvalues and eigenvectors of the dynamic matrix \cite{Erba2013}, which was obtained using the direct (frozen-phonon) method  \cite{crystal17,Pascale2004}.

\begin{figure}[t]
	\centering
	\includegraphics[width=0.7\textwidth]{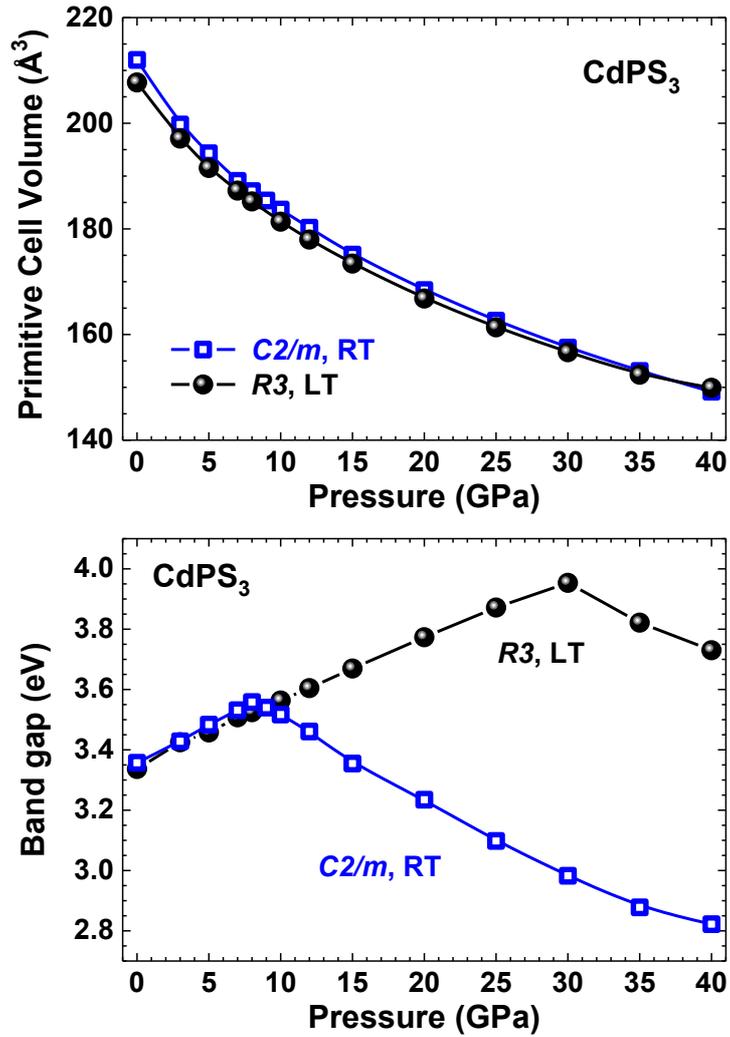}
	\caption{ Pressure dependence of the calculated primitive cell volume and the band gap $E_g$ 
      for the the low-temperature (LT) trigonal $R3$ and room-temperature (RT) monoclinic $C2/m$  CdPS$_3$ phases. }
	\label{fig5}
\end{figure}

Finally, the lattice parameters and atomic fractional coordinates were optimized at several selected pressures in the range of 0--40~GPa for the $R3$ and  $C2/m$ phases of CdPS$_3$. 
The structure optimization at the required pressure was performed using the approach developed in  \cite{Jackson2015}.
Total and projected onto the set of atomic orbitals density of states (DOS) at selected pressures are presented in Fig.\ \ref{fig4}. 
Pressure dependences of the primitive cell volume $V(P)$ and the band gap $E_g(P)$ are shown in Fig.\ \ref{fig5}.

\section{Results and discussion}\label{s:results}

The structural properties of bulk CdPS$_3$ from our LCAO calculations agree with the
experimental findings for both low- and room-temperature phases (Table\ \ref{table1}).  The obtained values of the Mulliken charges for three ions are close in both  phases and are equal to $Z$(Cd)=$+$0.61, $Z$(P)=$+$0.17 and $Z$(S)=$-$0.26. The indirect band gap was found 
in both phases (Fig.\ \ref{fig2}). Its value $E_g$=3.4~eV for the $C2/m$ phase is close to the experimental ones, ranging from 2.95 to 3.5~eV \cite{Fuentealba2020,Calareso1997,Brec1979,YANG2002,Du2016}.
The experimental data for the band gap in the $R3$ phase are not available to our knowledge, and
our calculations predict a value of 3.3~eV. 

The atomic displacement parameters (ADPs) $B_{\rm eq}$ (Table\ \ref{table1}) 
were used to estimate the amplitude of thermal vibrations of atoms in harmonic approximation.
The calculations were carried out at two temperatures (133~K for the $R3$ phase and 298~K for the $C2/m$ phase), for which the experimental data \cite{Ouvrard1985,BOUCHER1994,Boucher1995} are available.  Comparison of the calculated and experimental ADP values for Cd, P and S atoms shows that the theoretical values are systematically lower. Nevertheless, the relative ADP values (larger for Cd and smaller for P) are reproduced. The large values of the experimental ADPs for Cd atoms were attributed in the past to their apparent off-center displacement caused by a second-order Jahn-Teller
effect \cite{BOUCHER1994,Zhukov1995}.  

The total density of states (DOS) and its projection onto atoms (Fig.\ \ref{fig3}) and  onto a set of atomic orbitals (the s, p and d orbitals of Cd, P and S atoms)  (Fig.\ \ref{fig4}) allow us to draw a conclusion about the nature of electronic states in the valence and conduction bands.

As one can see, there is no big difference in DOSs for the $R3$ and  $C2/m$ phases at 0~GPa.  
The valence band maximum is dominated by 3p(S) orbitals with much lower contribution from 4d(Cd) and 3p(P) orbitals.  
The conduction band minimum is mainly comprised of the 3p(S), 3s(P) and 3p(P) orbitals with some admixture of the 5s(Cd) and 3s(S) orbitals.  The important role of the 3p(P) states at the bottom of the conduction band was demonstrated recently in FePS$_3$ \cite{Evarestov2020a}.   

An increase in pressure has a similar effect on the density of states in the $R3$ and $C2/m$ phases. The contribution of the 5p(Cd) orbitals appears at the bottom of the conduction band in addition to the contributions of the 3p(S), 3s(P), 3p(P), 5s(Cd) and 3s(S)  orbitals. At largest pressures, the edges of the valence and conduction bands broaden, which leads to a decrease in the band gap. 
 
The effect of pressure on the CdPS$_3$ structure is shown in Fig.\ \ref{fig5}. Compression leads to a gradual reduction in the volume  of the primitive cell in both phases, however, the pressure dependence of the band gap passes through a maximum, but at different pressures.  The band gap increases almost linearly in the $R3$ phase reaching a maximum value of $\sim$4~eV at $\sim$30~GPa and then decreases, while in the $C2/m$ phase, the maximum value of the band gap of $\sim$3.6~eV is observed at $\sim$8~GPa.   

Pressure dependence of the interatomic distances is summarized in Table\ \ref{table2} at three selected pressures.  The calculations show that upon compression all interatomic distances are reduced, but in different ways. In the $R3$ phase, Cd atoms coordinated octahedrally by S atoms displace along the $c$-axis direction  upon increasing pressure, so that the CdS$_6$ octahedra become more distorted at 30~GPa.  In the $C2/m$ phase, their displacement is much smaller and occurs along the $b$-axis direction, so that the CdS$_6$ octahedra are distorted less.

Different displacements of Cd atoms in the $R3$ and $C2/m$ phases have different effects on the Cd--Cd and Cd--P interatomic distances. In both phases, their mean values decrease, respectively, by 0.23-0.25~\AA\ and 0.25-0.30~\AA\ upon compression.  However, only in  the $C2/m$ phase there are two different Cd--Cd distances of about 3.55~\AA\ and 3.58~\AA\ at 0~GPa, and their difference increases at 30~GPa up to 0.23~\AA\ (the distances are 3.20~\AA\ and 3.43~\AA). The Cd--P distribution is split in both phases, and the splitting increases from 0.01-0.03~\AA\ at 0~GPa to 0.07-0.11~\AA\ at 30~GPa.  

The interatomic P--S distances are reduced by 0.07~\AA\  in both phases under compression from 0 to 30~GPa. Similar behaviour is observed for the P--P distances in the P$_2$S$_6$ group: they are reduced, respectively, by 0.11~\AA\ and 0.08~\AA\ in the $R3$ and $C2/m$ phases at 30~GPa.

\section{Conclusions}\label{s:conc}
A detailed investigation of the electronic and atomic structure of a bulk 2D layered van-der-Waals compound CdPS$_3$ was performed in the low-temperature ($R3$) and room-temperature ($C2/m$)  phases using first-principles LCAO calculations with hybrid M06 exchange-correlation functional.  The obtained results agree with the known experimental crystallographic parameters. 
Close indirect band gaps of 3.3~eV and 3.4~eV were found for the $R3$ and $C2/m$ phases at 0~GPa, respectively.
At the same time, the band gap has a pronounced dependence on pressure with a maximum value at pressures that differ 
in the two phases. Hence, pressure can be used to modulate the electronic structure of  bulk CdPS$_3$.

\section*{Acknowledgements}
A.K. is grateful to the Latvian Council of Science project no. lzp-2018/2-0353 for financial support.
Institute of Solid State Physics, University of Latvia as the Center of Excellence has received funding from the European Union's Horizon 2020 Framework Programme H2020-WIDESPREAD-01-2016-2017-TeamingPhase2 under grant agreement No. 739508, project CAMART2.

\bibliography{references}

\newpage


\begin{table}
	\footnotesize
	\caption{Crystallographic parameters, atomic displacement parameters (ADPs) $B_{\rm eq}$ (at 133~K for $R3$ and 298~K for $C2/m$ phases) and band gap $E_g$ values for CdPS$_3$ in low-temperature $R3$  and room-temperature $C2/m$ phases. }
	\label{table1}        
	\centering 
	\renewcommand{\arraystretch}{1.0}
	\begin{tabular}{lllll} 
		\\
		\hline 
		& \multicolumn{2}{c}{Space group $R3$ (146)}  &  \multicolumn{2}{c}{Space group $C2/m$ (12)} \\  
		
		& Experiment \protect\cite{Boucher1995} & LCAO    & Experiment \protect\cite{Ouvrard1985,BOUCHER1994} & LCAO      \\        
		\hline
		a (\AA)  & 6.224  & 6.189   & 6.218    & 6.167     \\
		b (\AA)  &        &   & 10.763   &  10.690          \\
		c (\AA)  & 19.490 & 18.784  & 6.867    & 6.752  \\
		
		$\beta$ ($^\circ$) &  &  & 107.58   & 107.72  \\
		& & & & \\
    	y(Cd1)  &  &  & 0.3322   & 0.3324   \\
		
		z(Cd1)  & 0.3291 & 0.3301 &    &    \\
		z(Cd2)  & 0.3380 & 0.3368  &    &    \\		
					& & & & \\	
		x(P1)   &          &  & 0.0551   & 0.0575  \\	
		z(P1)   & 0.3900  & 0.3930  	 & 0.1698    & 0.1741   \\
		z(P2)   & 0.2757  & 0.2739 	 &    &   \\
				& & & & \\
		x(S1)  & 0.3450  &0.3416  &0.7686  &  0.7618 \\
		y(S1)  & 0.3214  &0.3266   &   &   \\		
		z(S1)  & 0.4201  &0.4225 & 0.2541  & 0.2582   \\
			
		x(S2) & -0.0168 &-0.0083  & 0.2401   & 0.2467   \\
		y(S2) & 0.3422  & 0.3401  & 0.1567  & 0.1610   \\
		z(S2) & 0.2471  & 0.2443   & 0.2584  & 0.2628  \\ 
		& & & & \\		
        $B_{\rm eq}$(Cd) (\AA$^2$) & 0.856   & 0.31  & 1.84   & 0.67  \\		
		$B_{\rm eq}$(P) (\AA$^2$) &  0.54    & 0.19  & 0.91  &  0.34 \\
        $B_{\rm eq}$(S) (\AA$^2$) &  0.65    & 0.27  & 1.23, 1.28   & 0.52, 0.55  \\		
		
		& & & & \\		
		$E_g$ (eV) &      & 3.3  & 2.95~eV \protect\cite{Fuentealba2020}   & 3.4 \\
 			&      &   &  3.06 \protect\cite{Calareso1997}   & \\
 			 &      &  &  3.5 \protect\cite{Brec1979,YANG2002,Du2016}   & \\
		\hline
	\end{tabular}
\end{table}

\newpage 

\begin{table}
	\footnotesize
	\caption{Pressure dependence of the calculated interatomic distances in CdPS$_3$ in low-temperature  $R3$  and room-temperature $C2/m$ phases at three selected pressures. The difference $\Delta$ between the distances at 30 and 0~GPa is shown in the last column. }
	\label{table2}     
	\centering 
	\renewcommand{\arraystretch}{1.0}
	\begin{tabular}{lcccc} 
		\\
		\hline 
 &    0~GPa   &   8~GPa    &  30~GPa  & $\Delta$ \\
 		\hline
      \multicolumn{5}{c}{$R3$ phase} \\
Cd--S	 &  2.67	 &  2.60	 &  2.49 &  -0.18  \\
	 &  2.70	 &  2.65	 &  2.57 &  -0.13  \\
$\langle$Cd--S$\rangle$	 &  2.69	 &  2.63	 &  2.53 &  -0.16  \\
    & & & &  \\
P--S	 &  2.07	 &  2.05	 &  2.00 &  -0.06  \\
P--P     & 	2.24	 &  2.20	 &  2.13 &  -0.11 \\
Cd--Cd	 &  3.58	 &  3.49	 &  3.34 &  -0.23  \\
    & & & &  \\
Cd--P	 &  3.73	 &  3.62	 &  3.44 &  -0.28  \\
	     &  3.76	 &  3.69	 &  3.55 & 	-0.22  \\
$\langle$Cd--P$\rangle$	 &  3.75	 &  3.66	 &  3.50 & 	-0.25  \\
    & & & &  \\
      \multicolumn{5}{c}{$C2/m$ phase} \\
Cd--S	 &  2.67	 &  2.60	 &  2.50 &  -0.17  \\
	 &  2.68	 &  2.61	 &  2.51 &  -0.17  \\
	 &  2.70	 &  2.65	 &  2.54 &  -0.16  \\
$\langle$Cd--S$\rangle$	 &  2.68 &  2.62 &  2.52 &  -0.17  \\
& & & &  \\
P--S	 &  2.07	 &  2.04	 &  2.00 &  -0.07  \\
P--P     & 	2.24	 &  2.21	 &  2.16 &  -0.08 \\
& & & &  \\
Cd--Cd	 &  3.55	 &  3.43	 &  3.20 &  -0.36  \\
     	 &  3.58	 &  3.52	 &  3.43 & 	-0.15  \\
$\langle$Cd--Cd$\rangle$ &  3.57 &  3.47 &  3.31 & 	-0.25  \\
& & & &  \\
Cd--P	 &  3.73	 &  3.61	 &  3.41 &  -0.32  \\
     	 &  3.74	 &  3.64	 &  3.47 & 	-0.27  \\
$\langle$Cd--P$\rangle$	 &  3.73  &  3.63 &  3.44 & -0.30  \\

		\hline
\end{tabular}
\end{table}

\newpage

\end{document}